\documentclass[final,1p,times]{elsarticle} 
\usepackage{graphicx} 
\usepackage{amssymb} 
\usepackage{amsthm} 
\usepackage{lineno} 

\journal{Nuclear Physics A} 
\begin{document} 

\begin{frontmatter} 


\title{Stepping outside the neighborhood of $T_c$ at LHC}

\author{Urs Achim Wiedemann$^{a}$}

\address[a]{Physics Department, Theory Unit, CERN, 
CH-1211 Gen\`eve 23, Switzerland}

\begin{abstract} 
{\it "As you are well aware, many in the RHIC community are interested in the LHC
heavy-ion program, but have several questions: What can we learn at the
LHC that is qualitatively new?  Are collisions at LHC similar to RHIC
ones, just with a somewhat hotter/denser initial state?  If not, why
not?  These questions are asked in good faith, and this talk is an
opportunity to answer them directly to much of the RHIC community."}\\
With these words, the organizers of Quark Matter 2009 in Knoxville
invited me to discuss the physics opportunities for heavy ion collisions at the LHC
without recalling the standard arguments, which are mainly based on the extended
kinematic reach of the machine. In response, I emphasize here that lattice QCD 
indicates characteristic qualitative differences between thermal physics in the 
neighborhood of the critical temperature ($T_c < T < 400 - 500\, {\rm MeV}$) and thermal 
physics at higher temperatures ($ T > 400-500\, {\rm MeV}$), 
for which the relevant energy densities will be solely attainable at the LHC. 
\end{abstract} 

\end{frontmatter} 

\linenumbers 

\section{From AGS to SPS, from SPS to RHIC, from RHIC to LHC}

In the study of nucleus-nucleus collisions, an order of magnitude
increase in center of mass energy has always been accompanied 
by major discoveries. The move of heavy ion experiments in the late 1980s 
from the Brookhaven Alternative Gradient Synchrotron to the CERN SPS illustrates 
this as clearly as the move a decade later from the CERN SPS to experiments
at the relativistic heavy ion collider RHIC. In both cases, the increased kinematic
reach gave access to qualitatively novel characteristics of the collision system, 
such as the discovery of $J/\Psi$-suppression at the CERN SPS, or the 
discovery of leading hadron suppression ("jet quenching") at RHIC. Also,
in both cases, observations made already at the lower center of mass
energy could be characterized in much more detail and reached a more mature
understanding at higher center of mass energy. For instance, hadronic abundances
close to thermal equilibrium had been measured already at the AGS, but our
current understanding of hadrochemistry in terms of a grand canonical description
was firmly established only at the CERN SPS, where 
two-parameter fits of thermal models accounted for the abundances of a large 
number of hadronic resonances, including rare multi-strange baryon yields which 
showed up to a factor 20 enhancement. Another example is elliptic flow, which had been 
measured in great detail at the CERN SPS. But it was only the gentle but continuous 
increase of the
elliptic flow signal with center of mass energy, and its particle species dependence 
at transverse momenta inaccessible at the CERN SPS, which gave strong support
to the current hydrodynamic interpretation of elliptic flow at RHIC. Furthermore, the move to
a higher center of mass energy and the accompanying increase in precision
and/or kinematic reach repeatedly initiated novel developments in theory. 
For instance, our current understanding of signatures of chiral symmetry restoration
and thermal modifications of vector mesons has been largely developed and refined 
by the interplay of theory and experiment at the CERN SPS. The same can be said
about HBT two-particle correlations or the global picture of expansion dynamics in
the context of the SPS, or the theory of jet quenching and of saturation in the context of RHIC. 

These developments are likely to be continued at the LHC. Discoveries arise
as a consequence of (logarithmic) increases in 
kinematic reach and/or substantial improvement of instrumentation or precision. 
Understanding arises from a tight interplay between theory and experiment on the
newly accessible measurements. For the LHC heavy ion program, the most foreseeable 
advances due to kinematic reach and instrumentation have
been reiterated repeatedly in plenary talks. For instance, a factor $\sim 30$ increase in 
center of mass energy from RHIC to LHC will make many rare hard probes for the first
time measurable and it will provide abundant samples of jets up to $E_T \approx 200$
GeV. The increased kinematic reach will also add qualitatively novel insight to many
soft characteristics of heavy ion collisions: For instance, the abundance of charmed and 
beauty hadrons will provide novel tests for our understanding of the hadrochemical
composition in heavy ion collision. Extending the measurements of elliptic flow to 
systems with denser initial conditions has the potential of falsifying or refining our current 
interpretation in terms of fluid dynamic evolution. Also, a factor $\approx 30$
increase in center of mass energy tests particle production at unprecedented
low Bjorken-$x$, where saturation phenomena may leave characteristic traces. 
Beyond these and other foreseeable physics opportunities, there is the exciting 
thought that a factor 30 jump into the unexpected is always a unique chance for finding
the unexpected.

The novel physics opportunities, alluded to sketchily in the above paragraphs, could
be discussed in much more detail by extrapolating the established (SPS and RHIC)  phenomenology to the higher LHC center of mass energy, and discussing the resulting expectations in the light of the established experimental capabilities. 
However, the 
organizers discouraged me from reiterating once more these standard arguments. 
They rather encouraged me to reflect on the question of whether there are first principle
calculations in quantum field theory, which could support the idea that the matter
produced in heavy ion collisions at the LHC is qualitatively different from the matter 
produced at RHIC. In the following, I would like to recall some results from lattice QCD,
which may support a positive answer.

\begin{figure}[ht]
\centering
\includegraphics[scale=0.52]{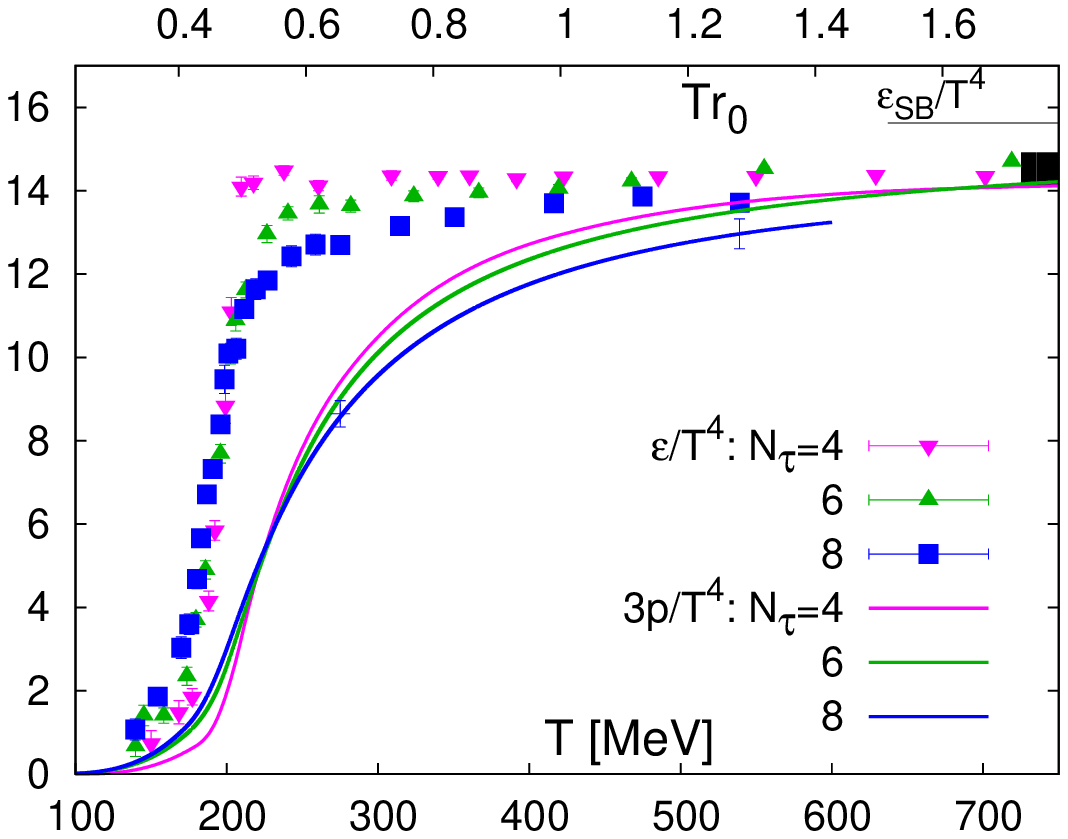}	
\includegraphics[scale=0.52]{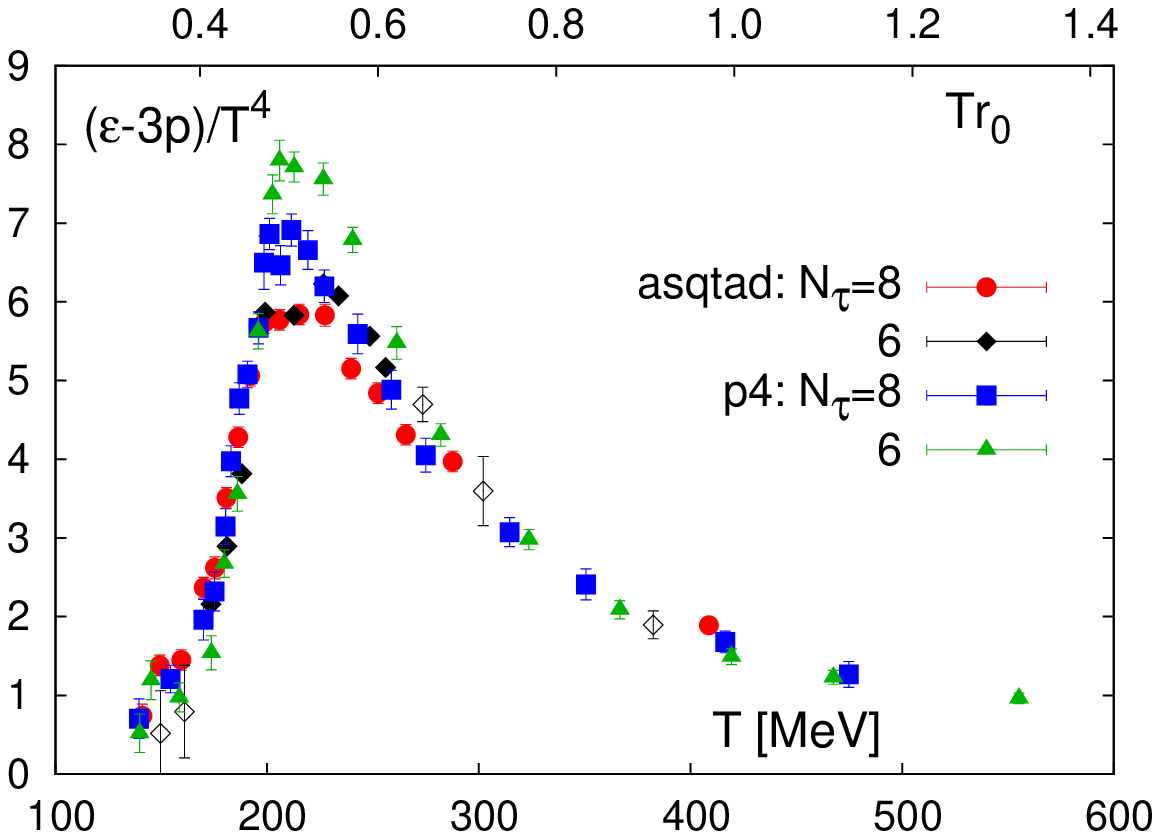}	       
\caption[]{(color online) Lattice data for finite temperature 2+1 flavor QCD with physical strange 
quark mass and almost physical light quark mass. Left hand side: Temperature
dependence of energy density $\epsilon$ and three times the pressure in units of 
$T^4$. Right hand side: The trace anomaly $(\epsilon - 3p)/T^4$. Data are for
lattices with different temporal extent $N_\tau$ and for different actions. Figure
taken from Ref.~\cite{Bazavov:2009zn}.}
\label{fig1}
\end{figure}
\section{QCD thermodynamics outside the neighborhood of $T_c$}

Most generally, heavy ion physics aims at understanding how collective 
phenomena and macroscopic properties of matter emerge from the fundamental
interactions of the non-abelian quantum field theory QCD. 
Heavy ion physics addresses this question at the
highest energies and densities attainable in the laboratory, where partons are
expected to be the relevant physical degrees of freedom, and where thus the connection 
of collective phenoma to the QCD lagrangian is most
direct. Arguably the most striking collective phenomenon predicted from first principle
calculations in QCD is the occurrence of the QCD phase transition to a deconfined state with 
restored chiral symmetry. This phase transition is seen clearly for instance in 
the ratio $\epsilon/T^4$ of energy density over temperature to the fourth power, 
which shows a rapid, 
almost step-function increase ("rapid cross over") at the critical temperature, indicative 
of a sudden change of the number of physical degrees of freedom, see the left hand 
side of Fig.~\ref{fig1}.  

A priori, it is unclear to what extent the systems created in heavy ion collisions are sufficiently
close to (local) thermal equilibrium, and to what extent other confounding effects in their 
dynamical evolution are sufficiently unimportant to provide detailed tests of the
QCD lattice equation of state. Part of this question remains to be decided in an interplay
between experiment and theory. Based on Bjorken's pocket formula, however, one may
estimate the initial energy density attainable in heavy ion collisions. Typical
numbers for the initial temperature attained
in heavy ion collisions lie in the range of
\begin{eqnarray}
 &&200 < T^{\rm SPS}_{\rm initial} < 300\, {\rm MeV}\, , \nonumber \\
 && 300 < T^{\rm RHIC}_{\rm initial} < 500\, {\rm MeV}\, , \nonumber \\
 && 500 < T^{\rm LHC}_{\rm initial} < 1000\,  {\rm MeV}\, . \nonumber
\end{eqnarray}
These numbers come with significant uncertainties~\cite{BraunMunzinger:2003zd}, in 
particular since Bjorken's estimate is proportional to the (collision energy dependent) 
time $\tau_i$, at which the system is assumed to reach kinetic equilibrium. Taken at face 
value, however, the equilibration temperatures estimated above suggest that the 
CERN SPS experimental program
was the first to gain access to the QCD high temperature phase, while
RHIC was the first machine to create systems which spend a significant time of their
dynamical evolution in this high temperature phase. Following this logic, the LHC will be
the first machine to create initial conditions with energy densities, which lie far away
from the neighborhood of the phase transition. Given that the ratio $\epsilon/T^4$
is apparently featureless above $T_c$, one may then rightly ask whether heavy ion
collisions at the LHC "may be similar to RHIC ones, just with a somewhat hotter/denser 
initial state". However, while the figure of $\epsilon/T^4$ is arguably an iconographic
representation of the QCD phase transition, its featurelessness above $T_c$ is not
shared by many other thermodynamic quantities.

One thermodynamic quantity, which is not featureless above $T_c$, is 
the so-called interaction measure or trace anomaly $(\epsilon - 3\, p)/T^4$. It
vanishes asymptotically for $T\gg T_c$, but it shows a characteristic and quantitatively
important peak in the region close to but above $T_c$, see right hand side of Fig.~\ref{fig1}.
In view of the estimated initial temperatures given above, one may argue that 
the RHIC machine seems to have access only to the part of the high energy phase,
where the interaction measure is sizeable.
In contrast, LHC will be the first machine to test the high temperature phase in the
region of almost vanishing interaction measure, where the equation of state approaches 
that of a free gas $\epsilon \approx 3\, p$ and where the system shows approximate
conformal invariance.  We now turn to arguments, which support the view that properties of
the medium change qualitatively for $T > 400 - 500\, {\rm MeV}$ and that a different
set of techniques may be applicable for the description of the QCD high temperature phase 
far outside the neighborhood of $T_c$.

\subsection{The Polyakov loop}

\begin{figure}[ht]
\centering
\includegraphics[scale=0.52]{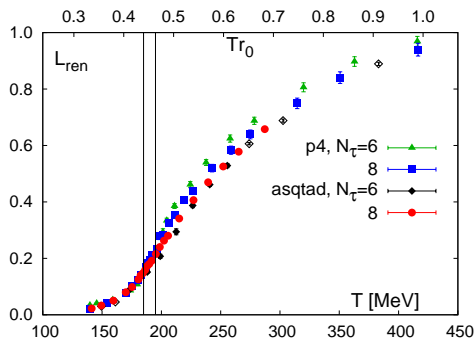}
\caption[]{(color online) The renormalized Polyakov loop $L_{\rm ren}$
of 2+1 flavor QCD 
for different lattice actions and lattices of different temporal extent . Figure
taken from Ref.~\cite{Bazavov:2009zn}.}
\label{fig2}
\end{figure}

For pure Yang-Mills theory without quarks, the trace of the Wilson line of the gauge field 
along the cyclic imaginary time direction
\begin{equation}
	{\rm Tr} L(\vec{x}) = {\rm Tr} \Bigg\lbrace {\cal P} 
	\exp \left[ i \int_0^\beta d\tau A_0(\tau,\vec{x})\right] \Big \rbrace
\end{equation}
is a good order parameter for deconfinement. This so-called Polyakov loop 
can be interpreted in terms of a static quark free energy,
\begin{equation}
	\langle {\rm Tr} L(\vec{x})\rangle  = 
	\exp \left[ - \beta \Delta F_q(\vec{x}) \right]\, .
\end{equation}
The Polyakov loop needs to be
renormalized in order to eliminate self-energy contributions to $\Delta F_q$. 
At low temperature, quarks are confined; as a consequence, the static quark free energy tends 
to infinity, and the thermal expectation value of the renormalized Polyakov loop 
$L_{\rm ren}$ vanishes. 
At high temperature, the system is deconfined, the static quark free energy becomes 
negligible, and $\langle {\rm Tr} L(\vec{x})\rangle$ approaches unity. These features are 
clearly seen in the lattice data of Fig.~\ref{fig2}.
%

\begin{figure}[ht]
\centering
 \includegraphics[width=1.0\textwidth]{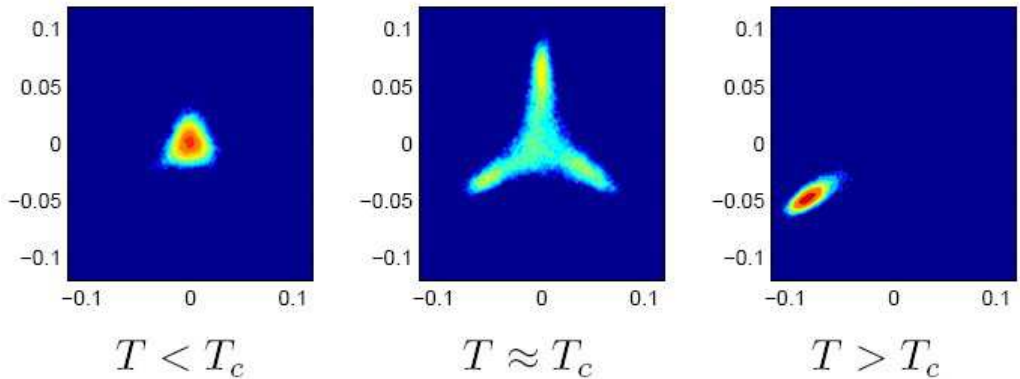}
\caption[]{(color online) Probability distribution of the real and imaginary part of ${\rm Tr}L$ 
for pure SU(3) Yang-Mills theory. In this lattice simulation, the modulus
$\vert  {\rm Tr}L\vert$ is not normalized to unity. The real part of ${\rm Tr}L$ 
is plotted along the $y$-axis. Figure is courtesy of A. Kurkela~\cite{Kurkela}. }
\label{fig3}
\end{figure}

Pure SU(3) Yang-Mills theory has an exact Z(3) center symmetry in its confined phase, 
which gets spontaneously broken at the deconfinement transition. The Polyakov loop
transforms under this exact $Z(3)$ symmetry,
${\rm Tr} L(\vec{x})  \to z\, {\rm Tr} L(\vec{x}) $, $z\in Z(3)$.
The corresponding symmetry transformation has to do with the boundary conditions of the 
allowed gauge transformations in the temporal direction, and its effects can be directly seen 
in the minima structure of the effective potential of the Wilson line~\cite{Gross:1980br}. 
In the deconfined phase, 
the potential has three minima, located at ${\rm Tr} L = \exp\left[ 2\pi\, i\, k/3\right]$, $k=0,1,2$.
All three minima are physically equivalent and the system arbitrarily chooses one. 
For temperatures below $T_c$, the $Z(3)$ symmetry is restored. Figure~\ref{fig3}
shows lattice results which illustrate this clearly. 

Adding quarks to SU(3) Yang-Mills theory breaks the center symmetry explicitly.
The three minima in the effective potential of the Wilson line are not degenerate
any more, but the minimum at ${\rm Tr}L = 1$ is favored over the complex minima
at $\exp \left[2\pi\, i / 3 \right]$ and $\exp \left[4\pi\, i / 3 \right]$. The latter correspond 
to metastable configurations. However, even with dynamical quarks, the qualitative
aspects regarding fluctuations between different minima are similar to the pure
gauge theory. As a consequence, the qualitative change of 
breaking the $Z(3)$ symmetry by adding quarks leads only to very mild numerical 
changes of the thermal expectation value $\langle {\rm Tr} L(\vec{x})\rangle$
(see e.g. Fig. 2 of Ref.~\cite{Petrov:2007ug}). These observations support the 
view that even in the presence of quarks the Polyakov loop can still serve as an 
indicator of the transition to the QCD high temperature phase.

\subsection{The Polyakov loop in the range $T_c < T < 3\, T_c$ and above}

We consider now in more detail the physics, which determines the shape of 
$\langle {\rm Tr} L(\vec{x})\rangle$.
At temperatures above $4 - 5\, T_c$, the system resides in one of the three 
$Z(3)$ minima and the renormalized Polyakov loop takes the asymptotic value of unity.
Lowering the temperature, one observes a qualitative change that starts to set in 
around $2 - 3\, T_c$, see Fig.~\ref{fig2}.  The physics reason is that fluctuations between 
the minima get more important, and thus the expectation value of the
Polyakov loop gets smaller since it averages over these minima.  Finally, as one lowers
the temperature to the critical one, the system is tunneling constantly between the different 
minima and it resides in different ones in different spatial regions. As a consequence,
the expectation value of the Polyakov loop vanishes.

In short, what drives this transition are fluctuations between different minima, and these
lead well above $T_c$ to a qualitative change in the characteristics of the QCD high 
temperature phase. We emphasize that this qualitative change occurs in a kinematic 
range of temperature, in which the QCD coupling constant varies only mildly. 
This illustrates that the mild logarithmic scale dependence of the strong coupling
constant above $T_c$ does not imply that major thermodynamic characteristics of 
QCD show a negligible evolution for temperatures above $T_c$. Rather, what matters
for the temperature dependence of the Polyakov loop is not the temperature dependence
of the coupling constant $\alpha_s(T)$, but rather the importance of fluctuations whose
dynamical relevance changes dramatically with the distance to the QCD phase 
transition. 

To sum up, the Polyakov loop illustrates that there is a qualitative change in the 
physics of the quark gluon plasma between a broad transition region
$T_c < T < 3\, T_c$ and temperatures higher than $3\, T_c$.  According to our 
simple estimates, the higher energies at which the renormalized Polyakov loop reaches 
its high-temperature limit $\langle {\rm Tr} L(\vec{x})\rangle \to 1$ will be 
accessible at the LHC but could not be explored yet at RHIC.

\subsection{The return of quasi-particle models?}
As discussed above, the $Z(3)$ center symmetry plays a crucial role in the region
$T_c < T < 3\, T_c$, but not for higher temperatures. Remarkably, in almost
all quasiparticle models \cite{Blaizot:2000fc,Kajantie:2000iz}, such as those
based on high-temperature dimensional reduction, one 
ends up explicitly breaking the center symmetry. These techniques
rely on expanding around a minimum field configuration, which resides in one of the
$Z(3)$ minima (For recent work aimed at circumventing this ansatz and working
with a $Z(3)$ symmetric effective theory, see 
Refs.~\cite{Vuorinen:2006nz,deForcrand:2008aw}.). In general, since quasi-particle
models expand around a single minimum without accounting for the other (almost) 
degenerate minima, such models cannot be expected to account
reliably for bulk thermodynamic quantities in a temperature range which is dominated
by the fluctuations between different $Z(3)$ minima. This field-theoretic observation
is in line with the general statement that strong coupling techniques are needed to 
describe thermodynamic properties in the temperature range accessible at RHIC. 

For temperatures above $3\, T_c$, this argument does not hold any more:
a compelling reason for why weak coupling techniques are inapplicable
is gone. This does not imply automatically that weak coupling techniques are 
applicable, but several observations indicate indeed that they may become
applicable for $T> 3\, T_c$, while they clearly fail at lower temperatures. For 
instance, it is a generic feature of weakly coupled quasi-particle models
that  the difference $\epsilon - 3\, p$ is proportional to $T^4$. In contrast to
this weak coupling expectation, lattice data indicate a quadratic 
dependence on $T$ in the range $T_c < T < 3\, T_c$~\cite{Cheng:2007jq}. 
For temperatures above $3\, T_c$, however, the trace anomaly vanishes
approximately, and this behavior is compatible with a weakly coupled
quasi-particle model. 

These observations may suggest that the "strong coupling paradigma", which has 
been developed in the context of RHIC phenomenology, may not extend to the entire
thermodynamical range accessible at the LHC. In other words: 
that the strong coupling constant 
changes only logarithmically on the scale between RHIC and LHC initial temperatures
should not be construed as implying the inapplicability of weak coupling techniques. 
There are field theoretic arguments, which indicate that weak coupling 
techniques can account for thermodynamic properties above $3\, T_c$ 
if not applied to the physics of elementary partonic quanta but to effective thermal 
degrees of freedom.  The kinematic reach of LHC opens the window for a novel
conceptual debate about how to view hot and dense matter far away from the 
neighborhood of $T_c$.

\section{Summary}
We have used Bjorken estimates for the initial temperature to relate lattice 
simulations of finite temperature field theory to the experimental conditions in
heavy ion collisions. This supported arguments that RHIC has explored 
an intermediate temperature range up to $T \approx 400 - 500\, {\rm MeV}$, 
in which the value of the Polyakov loop deviates significantly from unity and
where the interaction measure $\left(\epsilon - 3\, p \right)/T^4$ indicates strong
deviations from the equation of state of an ideal gas. These features are 
characteristically different from those of the genuine high temperature phase
of QCD, which sets in only for temperatures above 400 - 500 MeV, and which is 
thus only accessible by experiments at the LHC. At face value, these lattice
data hence suggest that the matter produced in heavy ion collisions at the LHC
may be characteristically different from the matter produced at RHIC. 

In particular, the field theoretic motivation of quasi-particle models relies 
on expanding around minimum field configurations, which reside in one
of the $Z(3)$ minima. As a consequence, such quasi-particle models are 
unlikely to capture the bulk properties of thermal QCD in a temperature
range up to 400 - 500 MeV, which is dominated by fluctuations between
the $Z(3)$ minima. On the other hand, both the approximate validity of
an ideal equation of state for $T > 400 - 500\, {\rm MeV}$, and the value
of the Polyakov loop in this temperature range supports the idea that
a weak coupling description of the medium in terms of some effective 
physical degrees of freedom becomes applicable. In other words, the 
question of whether a perturbative description of the medium produced
at LHC is applicable may not depend so much on the value of the strong
coupling constant, which changes only logarithmically. It may rather depend
on the availability of a stable minimum field configuration, on top of which 
a perturbative expansion in effective degrees of freedom can be based. 
Since the value of the Polyakov loop changes dramatically from the RHIC
to the LHC energy range, lattice QCD strongly supports the view that such a stable
minimum field configuration exists only for temperatures reachable at the
LHC, while lower temperatures are dominated by fluctuations between
different metastable minima. 

Relating first principle calculations of lattice QCD to the phenomenology of
heavy ion collisions is known to involve significant uncertainties. In particular,
it relies on controlling possibly confounding factors. These arise for instance from
non-equilibrium physics. Within a thermalized evolution, they can also arise 
from the known strong collective dynamics, which turns essentially all medium effects into
averages over different periods of the expansion (and, a fortiori, into averages
over different temperatures). Heavy ion collisions at the LHC will be initialized
at much higher energy densities than those at RHIC, but they will live through
the entire range of energy density explored at RHIC. The above qualitative arguments 
allow us to state that there are arguments from 1st principle calculations
in lattice QCD, which indicate that the matter produced at the LHC should not be 
assumed to be solely somewhat hotter or denser than that produced at RHIC.
Rather, the analysis of data at the LHC may require a critical assessment of
the fundamental question of whether bulk thermodynamic quantities at the LHC
are best described by strong coupling techniques.


\section*{Acknowledgments} 
In developing the above arguments, I have profited substantially from
discussions with A. Vuorinen and exchanges  with A. Kurkela. 

\end{document}